\documentclass[9pt,twocolumn,twoside]{pnas-new}
\usepackage{subfigure}
\usepackage{multirow}
\usepackage{amsmath}
\usepackage{mathrsfs}
\usepackage{chemfig}
\usepackage{mathtools}
\usepackage{relsize}
\usepackage{color}

\templatetype{pnasresearcharticle} % Choose template 
\title{Entropy-Controlled Cross-Linking in  Linker-Mediated Vitrimers}

\author[a,1]{Qun-Li Lei}
\author[a,b,1]{Xiuyang Xia}
\author[c]{Juan Yang}
\author[b]{Massimo Pica Ciamarra}
\author[a,2]{Ran Ni}

\affil[a]{School of Chemical and Biomedical Engineering, Nanyang Technological University, 62 Nanyang Drive, 637459, Singapore}
\affil[b]{Division of Physics and Applied Physics, School of Physical and Mathematical Sciences, Nanyang Technological University, 21 Nanyang Link, 637371 Singapore}
\affil[c]{Department of Chemistry, National University of Singapore, 117546 Singapore}

\leadauthor{Qun-Li Lei} 

\significancestatement{The recently developed linker-mediated vitrimers based on metathesis reactions offer new possibilities of processing cross-linked polymers with high mechanical performance in industry, while the design principle remains unknown.
Here we propose a theoretical framework for describing the system of linker-mediated vitrimers, in which entropy is found to play a dictating role. Our mean field theory agrees quantitatively with computer simulations, and provides guidelines for the rational design of the linker-mediated vitrimers with desired properties.
}

\authorcontributions{Author contributions: R.N. designed and directed the research; Q.-L.L. formulated the mean field theory; X.X. performed the computer simulations;
and all authors discussed the results and wrote the manuscript.}
\authordeclaration{The authors declare no conflict of interest.}
\equalauthors{\textsuperscript{1}Q.-L.L. and X.X. contributed equally to this work.}
\correspondingauthor{\textsuperscript{2}To whom correspondence should be addressed. E-mail: r.ni@ntu.edu.sg}


\begin{abstract}
Recently developed linker-mediated vitrimers based on metathesis of dioxaborolanes with various commercially available polymers have shown both good processability  and outstanding performance, such as mechanical, thermal, and chemical resistance, suggesting new ways of processing cross-linked polymers in industry, of which the design principle remains unknown [M. R{\"o}ttger, \emph{et al.}, \emph{Science} 356, 62 (2017)]. Here we formulate a theoretical framework to elucidate the phase behaviour of the linker-mediated vitrimers, in which entropy plays a governing role. We find that with increasing the linker concentration, vitrimers undergo a reentrant gel-sol transition, which explains a recent experiment [S. Wu, H. Yang, S. Huang, Q. Chen, \emph{Macromolecules} 53, 1180 (2020)]. More intriguingly, at the low temperature limit, the linker concentration still determines the cross-linking degree of the vitrimers, which originates from the competition between the conformational entropy of polymers and the translational entropy of linkers. Our theoretical predictions agree quantitatively with computer simulations, and offer guidelines in understanding and controlling the properties of this newly developed vitrimer system.
\end{abstract}

\dates{This manuscript was compiled on \today}
\doi{\url{www.pnas.org/cgi/doi/10.1073/pnas.XXXXXXXXXX}}

\begin{document}

\maketitle
\thispagestyle{firststyle}
\ifthenelse{\boolean{shortarticle}}{\ifthenelse{\boolean{singlecolumn}}{\abscontentformatted}{\abscontent}}{}

% If your first paragraph (i.e. with the \dropcap) contains a list environment (quote, quotation, theorem, definition, enumerate, itemize...), the line after the list may have some extra indentation. If this is the case, add \parshape=0 to the end of the list environment.

\dropcap{V}itrimers, { a new type of polymeric materials, are known to exhibit unique properties that combine the advantages of thermosets and thermoplastics. To be specific, they are mechanically robust and insoluble while also recyclable and malleable~\cite{montarnal2011silica,capelot2012metal}.
At low temperatures, vitrimers behave as cross-linked thermosets; while at high temperatures, the exchangeable bonds in the polymer network swap reversibly by the thermally-triggered reactions so that they behave as viscoelastic liquids~\cite{smallenburg2013patchy,ciarella2018dynamics,
ciarella2019understanding, meng2019elasticity,tito2019harnessing,wu2019dynamics,Denissen2016,Winne2019,zee2020,zhang2018polymer}.}
In the past decade, various chemical reactions, such as the transesterification reaction~\cite{capelot2012metal,brutman2014polylactide}, transamination reaction~\cite{denissen2015vinylogous,taynton2014heat}, alkoxyamine exchange reaction~\cite{wojtecki2011using,zhang2014application}, olefin metathesis~\cite{lu2012making}, thiol-disulfide exchange~\cite{pepels2013self}, have been used in the production of vitrimers for different applications.
However, the reactants or catalysts involved in reactions of the conventional vitrimers are usually not thermally or oxidatively stable, which is particularly detrimental for using the same equipments and conditions of processing thermoplastics~\cite{Denissen2016}.

Recently, a new type of linker-mediated vitrimers was developed based on the metathesis reaction of dioxaborolanes, in which the functionalized polymers with pendant dioxaborolane units react with bis-dioxaborolanes (cross-linkers), and the metathesis reaction here both cross-links the polymers and dynamically changes the polymer network~\cite{rottger2017high}.
The linker-mediated vitrimers have superior chemical resistance and dimensional stability, without the need of a catalyst, and can be processed like thermoplastics~\cite{garcia2017future}.
In this work, we propose a mean field theory combined with coarse-grained computer simulations to study the linker-mediated vitrimer system, and our results show that the entropy of free linkers plays a nontrivial role. We find that with increasing the concentration of free linkers, the vitrimer system undergoes a reentrant gel-sol transition, which was observed in a recent experiment~\cite{wu2020relationship}.
More interestingly, even at the low temperature limit, the cross-linking degree of vitrimers still depends on the concentration of free linkers, which essentially offers an extra degree of freedom in controlling the mechanical property of the resulting materials.

\section*{Results}

\begin{figure}[htbp]
\centering
\begin{tabular}{c}
        \resizebox{85mm}{!}{\includegraphics[trim=0.0in 0.0in 0.0in 0.0in]{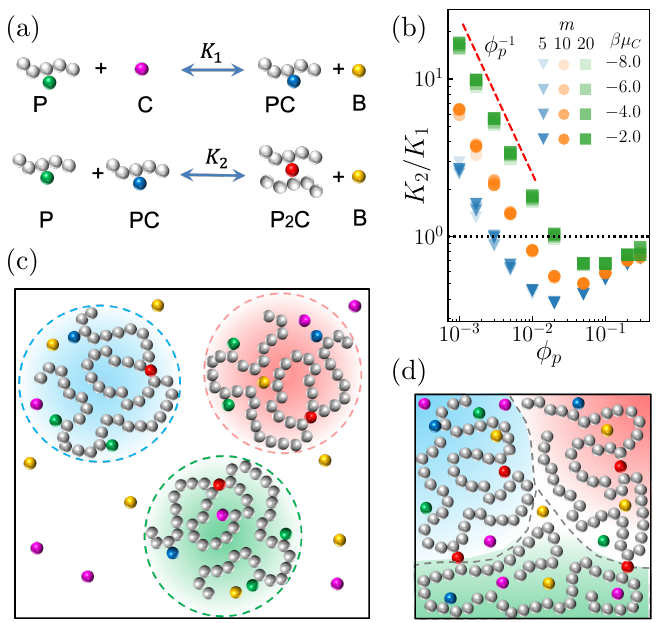} }
\end{tabular}
\caption{ \textbf{Vitrimer model.} (a): Illustration of the two step metathesis reactions in the vitrimer system. (b): $K_2/K_1$ as a function of $\phi_p$ for different $m$ and $\mu_{\rm C}$ at $n=100$ and $\beta \mu_{\rm B} = -3$. (c,d): Illustration of the heterogeneous dilute (c) and homogeneous dense (d) systems of the vitrimers.}
\label{Fig1}
\end{figure}

\subsection*{{ Coarse-Grained Model of Vitrimer}}
We consider a system of volume $V$ consisting of $N_{ poly}$ polymer chains, in which each polymer comprises of $n$ hard spheres of diameter $\sigma$.
As shown in Fig.~\ref{Fig1}a, on each polymer there are $m \gg 1$ precursors (reactive sites) P uniformly distributed, which can react with a cross-linker molecule C by forming a dangling PC bond and producing a byproduct free molecule B through metathesis reactions. Moreover, a dangling PC bond can further react with another intact precursor P to form a cross-linking $\rm P_2C$  bond and producing an additional free B molecule, and each cross-linker can form at most two bonds with two different precursors.
The metathesis reactions are reversible, and $\Delta G$ is the reaction energy.
$N_{\rm C}$ and $N_{\rm B}$ are the numbers of cross-linker molecule C and the byproduct molecule B in the system, which are controlled by the chemical potentials $\mu_{\rm C}$ and $\mu_{\rm B}$, respectively.
We define $n_{\rm PC}$ and $n_{\rm P_2C}$ as the average numbers of PC bonds with dangling cross-linkers and cross-linking P$_2$C  bonds per polymer, respectively.
Similarly, $n_{\rm P}$ is the average number of precursors per polymer that remain intact, and $N_i= N_{ poly} n_i $ is the total number of precursors or bonds of type $ i=\rm P,PC, P_2C$ in the system. The packing fraction of polymers is $\phi_p = \frac{\pi}{6} N_{poly} n \sigma^3$, and B and C are both modelled as hard spheres of diameter $\sigma$. 

{ We employ Monte Carlo (MC) simulations to investigate this coarse-grained hard-sphere-chain system.  An infinitely deep square-well tethered bond potential~\cite{nihspolymer} below is used to mimic the connectivity among covalently bonded polymer beads and cross-linkers,
\begin{equation}
V_\mathrm{bonds} (\mathbf{r},\mathbf{r}')=\left\{\begin{array}{ll}
0 & \left|\mathbf{r}-\mathbf{r}^{\prime}\right|<r_\mathrm{cut} \\
\infty & \text { else }
\end{array}\right.,
\end{equation}
where $r_\mathrm{cut}$ is the cutoff distance of the tethered bond, and we use $r_{\rm cut} = 1.5 \sigma$ throughout all simulations. See {\emph Materials and Methods} for the simulation details.}

\begin{figure*}[htbp]
\centering
\begin{tabular}{c}
        \resizebox{170mm}{!}{\includegraphics[trim=0.0in 0.0in 0.0in 0.0in]{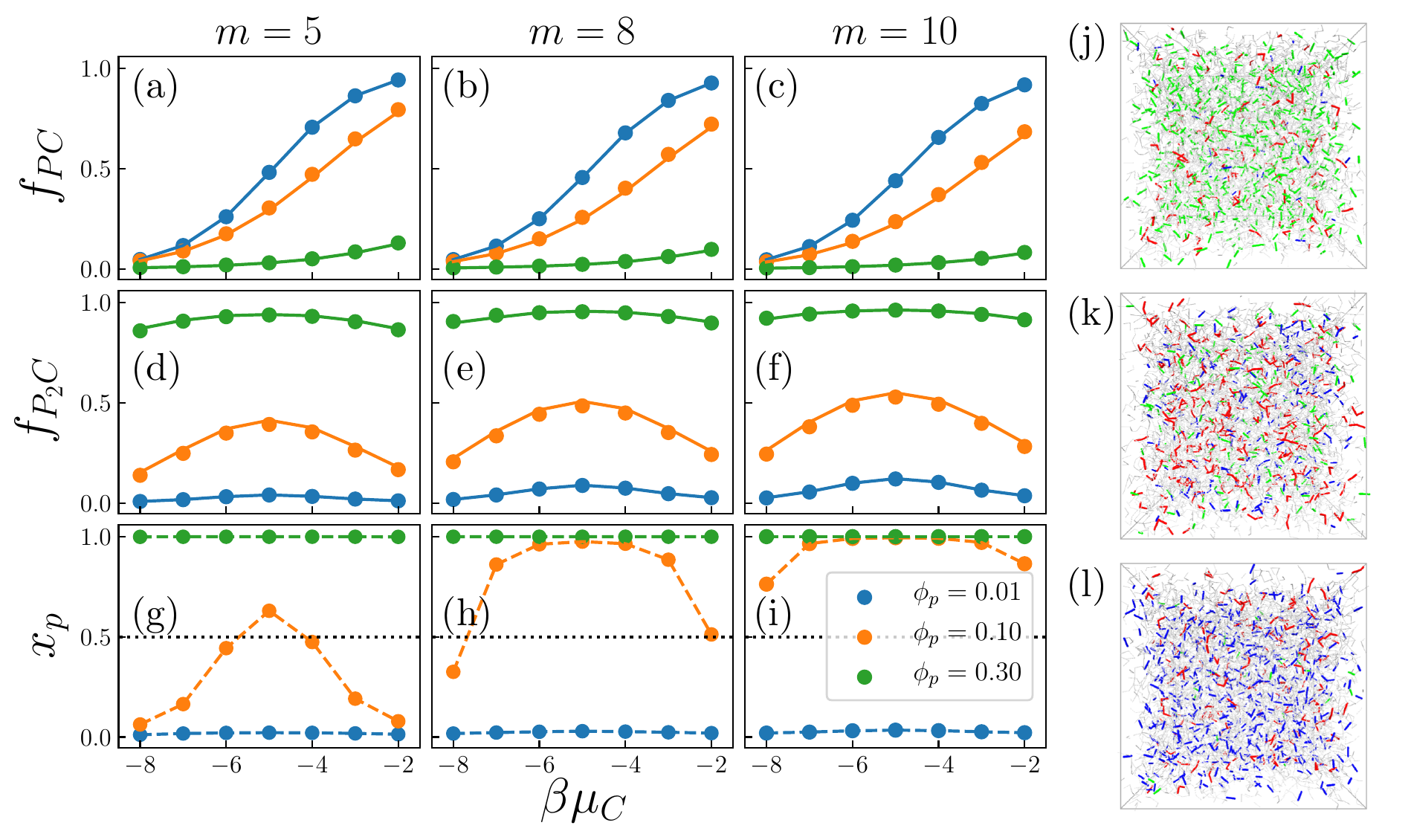} }
\end{tabular}
\caption{\textbf{Reentrant gel-sol transition.} (a-i): $f_{\rm PC},~f_{\rm P_2C},~x_p$ as a function of $\mu_{\rm C}$ for different $m$ and $\phi_p$. The solid lines in (a-f) are the theoretical predictions of Eqs.~\ref{f_P2C}, \ref{f_PC} and \ref{cform} { using $K_2/K_1$ obtained from simulations,  while the dashed lines in (g-i) are guide for eye. }. (j-l): Simulation snapshots for systems with $m=10$ and $\phi_p = 0.1$ at $\beta \mu_{\rm C}=-8.0$ (j), $-5.0$ (k) and $-2.0$ (l). Different colors represent different type of bonds based on Fig.~\ref{Fig1}a,
i.e., $\rm PC$, $\rm P_2C$, P and polymer backbone bonds are in blue, red, green and grey, respectively. Free B and C molecules are not drawn.
 In all simulations, $\beta \mu_{\rm B}=-3.0$, $\beta \Delta G=-2.0$, and $n=100$. }
\label{Fig2}
\end{figure*}

\subsection*{ Mean field theory  }
In the dilute limit (Fig.~\ref{Fig1}c), polymer chains are isolated. The distribution of reactive sites in the system is heterogeneous and the reactive sites can be seen as ideal gases confined in individual polymer blobs of volume $V_p=R_g^3$ with $R_g$ the radius of gyration of the polymer. Treating the polymer backbones implicitly as the crowding background, the free energy of the system can be written as
\begin{eqnarray} 
{\beta F} &=&   \sum\limits_{i=poly,\rm B,C}N_{i}{ \left[\ln \left( \frac{N_{i} \Lambda^3}{V}\right) -1\right]}  \notag \\
&& +  \sum_{i=\rm P,PC,P_2C} N_{i} \left[\ln \left( \frac{n_{i}\Lambda^3}{V_p} \right) -1\right]  + \beta F^{ex}_{\rm HS} \notag \\
&& -N_{\rm P_2C}k_B^{-1}  \Delta S -\beta \left(N_{\rm PC}+ {N_{\rm P_2C}} + N_{\rm C} \right) \mu_{\rm C}
 \notag \\
&&  + \beta (N_{\rm PC} +2  N_{\rm P_2C} ) \left(\Delta G +\mu_{\rm B} \right) - \beta N_{\rm B}\mu_{\rm B}, \label{free_energy}
\end{eqnarray}
where the first summation is on the ideal gas terms of polymer chains, free cross-linkers and byproduct molecules, and the second summation is on the ideal gas terms of different reactive sites in polymer blobs.
Here $\beta = 1/k_BT$ with $k_B$ and $T$ the Boltzmann constant and the temperature of the system, respectively, and $\Lambda$ is the de Broglie wavelength.
This ideal gas approximation of reactive sites offers a simple estimation of  the conformational entropy change during the formation of cross-linking $\rm P_2C$  bonds, for which we introduce $\Delta S$  as the entropy correction per cross-linking bond  { for any inaccuracy arising from the ideal gas approximation}. $ F^{ex}_{\rm HS} $ is the excess free energy based on Carnahan-Starling hard-sphere equation of state~\cite{hansen1990theory} arising from the excluded volume interaction, which accounts for the crowding effect in the system. The last three terms arise from the { bond formation in} metathesis reactions { (related with $\Delta G$)}, and the exchange of molecules with reservoir { (related with $\mu_{\rm B}$, $\mu_{\rm C}$)}. As the system approaches the dense regime (Fig.~\ref{Fig1}d), polymer blobs  begin to overlap and the distribution of reactive site become homogeneous. Thus, $V_p$ can be replaced by the available volume per polymer, i.e., $V_p=V/N_{poly}$.

We define the cross-linking degree of system as $f_{\rm P_2C}=2N_{\rm P_2C}/(N_{poly} m)$, i.e., the fraction of reactive sites that are cross-linked, and $f_i=N_i/(N_{poly} m),~(i= \rm PC, B, C)$. Using the saddle-point approximation, $ {\partial F}/{ \partial f_i }=0,~~(i=\rm PC, P_2C, B, C)$, one can obtain the equilibrium $f_{\rm PC}$ and $f_{\rm P_2C}$ as
\begin{eqnarray} 
f_{\rm P_2C}&=&\frac{\left[-(1+a) +\sqrt{(1+a)^2+4c}\right]^2}{4c  }, \label{f_P2C}  \\
f_{\rm PC}&=&\frac{-(1+a) +\sqrt{(1+a)^2+4c}}{2c  }, \label{f_PC}
\end{eqnarray}
with
\begin{eqnarray} 
a&=&  e^{\beta (\Delta G + \mu_{\rm B} - \mu_{\rm C})},  \label{parameter_a} \\
c&=&  { \frac{ 2 m \Lambda^3 }{V_p}  } e^{-\beta{\mu_{\rm C}} + k_B^{-1} \Delta S+ \beta \mu_{\rm  HS}^{ex} }, \label{parameter_c}
\end{eqnarray}
where $\mu^{ex}_{\rm HS}$ is the excess chemical potential originating from $F^{ex}_{\rm HS}$.  Here we note that the effect of $\mu_{\rm B}$ is the same as $\Delta G$.  Since $\mu^{ex}_{\rm HS}$ is a function of the packing fraction of the system, which also depends on $f_i$, { self-consistent iterations are needed to obtain $f_{\rm PC}$, $f_{\rm P_2C}$ and the equilibrium packing fraction of the system (see SI Appendix for details).}

\subsection*{Reaction equilibrium}
As shown in  Fig.~\ref{Fig1}a, the formation a cross-linking $\rm P_2C$  bond requires two metathesis reactions, with $K_1$ and $K_2$ the corresponding reaction constants, which satisfy
\begin{eqnarray} 
K_1  \rho_{\rm  P}  \rho_{\rm  C}  &=& \rho_{\rm  PC}  \rho_{\rm  B}, \label{K1} \\
K_2 \rho_{\rm P}  \rho_{\rm  PC} &=& \rho_{\rm  P_2C}  \rho_{\rm  B},  \label{K2} 
\end{eqnarray}
where  $\rho_i=N_i/V$ is the density of  specie $i$ in the system. In Ref.~\cite{rottger2017high,wu2020relationship}, the metathesis reactions in the first and second steps are of the same type, and one might think that the two reaction constants $K_2$ and $K_1$ should be equal. However, a recent experiment indicates that $K_2/K_1$ calculated from Eqs.~\ref{K1} and \ref{K2} can be significantly larger than 1~\cite{wu2020relationship}. To understand this, we first use the fact that in chemical equilibrium, the chemical potentials for all species satisfy
\begin{eqnarray} 
\mu_{\rm P} +\mu_{\rm C} & = &  \mu_{\rm PC} +\mu_{\rm B},  \label{reaction_eq_A} \\  
\mu_{\rm  P} +\mu_{\rm PC} & = & \mu_{\rm P_2C} +\mu_{\rm B},  \label{reaction_eq_B}
\end{eqnarray}
where $\mu_i$ ($i=\rm  P,~PC,~P_2C$) can be obtained by taking a direct derivative of Eq.~\ref{free_energy} with respect to $N_i$ { (see SI Appendix)}. Combining Eqs.~\ref{K1} and \ref{K2} with Eqs.~\ref{reaction_eq_A} and \ref{reaction_eq_B} we  have
\begin{eqnarray} \label{k2k1}
\frac{K_2}{K_1} &=&  \frac{V}{N_{poly} V_p} e^{ k_B^{-1} \Delta S} \nonumber \\
&  = &
\left\{
\renewcommand{\arraystretch}{2.0}
\begin{array}{lr}
  \mathlarger{ \frac{n v_0}{\phi_p R_g^3}} e^{ k_B^{-1}  \Delta S }   & ( R_g^3 \ll  V/N_{poly}   )\\
e^{ k_B^{-1}  \Delta S }  & (R_g^3 \gg  V/N_{poly} )
\end{array}
\right.
\end{eqnarray}
where $v_0 = \frac{\pi}{6}\sigma^3$ is the volume of a single bead on the polymer chain.
Eq.~~\ref{k2k1} predicts a powerlaw scaling $K_2/K_1\sim \phi_p^{-1}$ in the heterogeneous dilute regime, which saturates to $e^{ k_B^{-1}  \Delta S }$ in the homogeneous dense regime. In Fig.~\ref{Fig1}b, we plot $K_2/K_1$ measured in simulations as a function of $\phi_p$ for different $m$ and $\mu_{\rm C}$ with $n=100$.
{ One can see that $K_2/K_1$ approaches 1 corresponding to  $\Delta S \rightarrow 0$ in the homogeneous dense regime, which suggests that the ideal gas approximation for the reactive sites is quantitatively correct in this regime. In the heterogeneous dilute regime, the powerlaw scaling is also confirmed in Fig~\ref{Fig1}b.} In this regime, $R_g$ decreases with increasing $m$ due to more cross-linking within each polymer blob, which also raises $K_2/K_1$ as shown in Fig.~\ref{Fig1}b. Therefore, the observation of $K_2/K_1>1$ in experiments could
result from the inhomogeneous distribution of
reactive sites in the system~\cite{ricarte2018phase,wu2020relationship,Ralm2020ma}.

With the introduction of $K_2/K_1$, Eq.~\ref{parameter_c} can be re-written as
\begin{eqnarray} \label{cform}
c&=&{ \left( \frac{K_2}{K_1} \right) \frac{ 2 N_{poly} m \Lambda^3 }{V }  } e^{\beta({\mu_{\rm HS}^{ex}}  - \mu_{\rm C}) }, 
\end{eqnarray}
which does not explicitly depend on $V_p$ or $\Delta  S$. { Therefore, the prediction of $f_{\rm  P_2C}$ and $f_{\rm  PC}$ from our theory (Eqs.~\ref{f_P2C} and \ref{f_PC}) only requires $K_2/K_1$  which can be measured both in experiments and simulations, without invoking any fitting parameters like $V_p$ or $\Delta  S$}. In Fig.~\ref{Fig2}a-f, we plot $f_{\rm  P_2C}$ and $f_{\rm PC}$ as functions of $\mu_{\rm C}$ obtained from the simulations for systems of various $m$ and $\phi_p$ with $n = 100$ and $\beta \mu_{\rm B} = -3$.  { The theoretical results based on Eqs.~\ref{f_P2C}, \ref{f_PC} and \ref{cform} using the measured $K_2/K_1$ in simulations are shown as solid lines, which agree quantitatively  with the simulation results in a wide range of $\phi_p$ and $m$. The equilibrium packing fractions of the system $\phi_{\rm eq}$ obtained from self-consistent iterations also match well with the simulation results (see Table S1 in the SI Appendix).}
This suggests that our theoretical framework is robust and capable of quantitatively predicting the phase behaviour of the vitrimers in both heterogeneous dilute and homogeneous dense regimes.

\subsection*{Reentrant gel-sol transition}

Moreover, Eq.~\ref{f_P2C} has the limiting behaviour:
\begin{eqnarray} 
f_{\rm P_2C}(\mu_{\rm C}\rightarrow - \infty) &\simeq&  \frac{c}{a^2} \rightarrow 0, \\
f_{\rm P_2C}(\mu_{\rm C}\rightarrow + \infty) &\simeq& c \rightarrow 0.
\end{eqnarray}
One can also prove that $f_{\rm P_2C}$ reaches the maximum at  $\mu_{\rm C}= \mu_{\rm B} + \Delta G$ (see SI Appendix), and this implies that the cross-linking degree first increases and then decreases with increasing the concentration of cross-linkers, which quantitatively agrees with simulations as shown in Fig.~\ref{Fig2}d-f. This non-monotonic dependence of $f_{\rm P_2C}$ on the concentration of cross-linkers was also observed in a recent experiment~\cite{wu2020relationship}, { which was explained based on reaction equilibrium and mass conservations. Here our theory shows that} the counter-intuitive decrease of $f_{\rm P_2C}$ with increasing $\mu_{\rm C}$ at high cross-linker concentration is a purely entropic effect.
At high $\mu_{\rm C}$ limit, the reactive sites are fully bonded with cross-linkers forming either $\rm PC$ or $\rm P_2C$ bonds. Changing one $\rm P_2C$ bond to two dangling $\rm PC$ bonds hardly changes the energy of the system but increases the conformational entropy of polymer chains. Therefore, forming dangling $\rm PC$ bonds is more entropically favoured than forming cross-linking $\rm P_2C$  bonds, which drives the cross-linking degree towards zero at high $\mu_{\rm C}$.
This non-monotonic dependence of $f_{\rm P_2C}$ on $\mu_{\rm C}$ provides an additional axis in controlling the mechanical properties of vitrimers. As the percolation of the connected polymer cluster is an important indicator of the sol-gel transition of the system, we plot the fraction of polymers in the largest cluster $x_p$ as a function of $\mu_{\rm C}$ for various $m$ and $\phi_p$ in Fig~\ref{Fig2}g-i.
Generally, gelation occurs at $x_p \simeq 0.5$~\cite{tanaka2011polymer}, and as shown in Fig~\ref{Fig2}g-i, $x_p$ changes non-monotonically with increasing $\mu_{\rm C}$, for which the typical snapshots of the system are shown in Fig.~\ref{Fig2}j-l.
This implies a reentrant gel-sol transition in this linker-mediated vitrimer system.

\begin{figure}[ht]
\centering
\begin{tabular}{c}
        \resizebox{85mm}{!}{\includegraphics[trim=0.0in 0.0in 0.0in 0.0in]{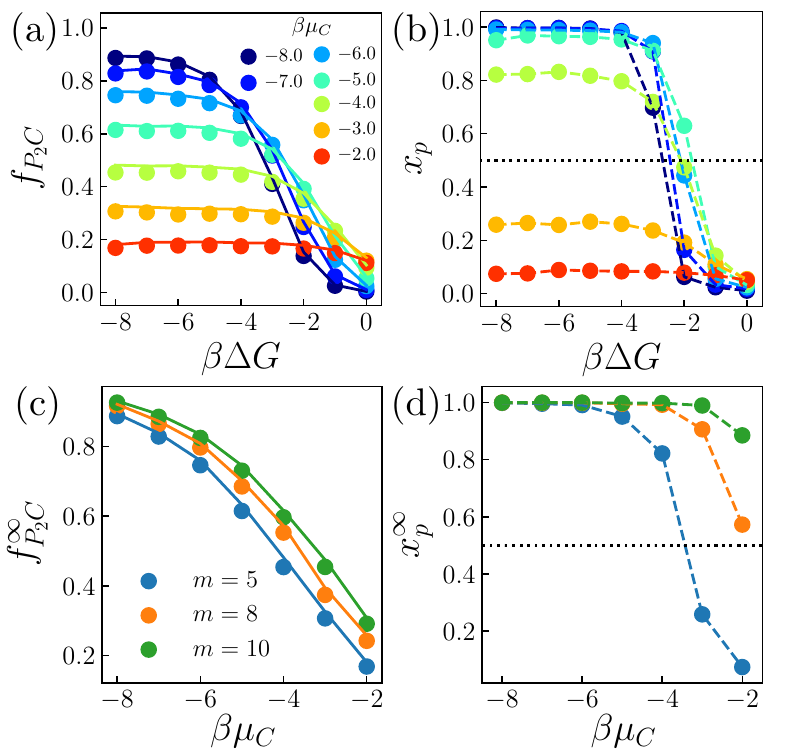} }
\end{tabular}
\caption{ \textbf{Entropy-driven cross-linking at low temperature.} (a-b): $f_{\rm P_2C}$ and $x_p$  as a function of reaction energy $\Delta G$ for different $\mu_{\rm C}$ at $m=5$, where the solid lines in (a) are the theoretical predictions of Eq.~\ref{f_P2C}. (c-d): Saturated $f^{\infty}_{\rm P_2C}$  and $x_p^{\infty}$ as a function of $\mu_{\rm C}$ for different $m$ at $\beta \Delta G=-8.0$, where the solid lines in (c) are the theoretical predictions of Eq.~\ref{finf}. { In (b,d), the dashed lines are guide for eye.} In all simulations, $\beta \mu_B=-3.0$, $n=100$, and $\phi_p = 0.1$. }
\label{Fig3}
\end{figure}

\subsection*{Entropic effects at low temperature}
Another interesting theoretical prediction is that in the limit of $\beta \Delta G \rightarrow -\infty$, one has  $f_{\rm P_2C} +f_{\rm PC} \simeq 1 $ and
\begin{eqnarray} \label{finf}
 {f_{\rm P_2C}}  &\simeq &   1 - \frac{ \sqrt{1+4c} -1 }{2c} < 1,
\end{eqnarray}
which implies that at the low temperature limit ($\beta \Delta G  \rightarrow -\infty$) or the dilute limit of byproduct molecules ($\beta \mu_{\rm B} \rightarrow - \infty$), the system is not fully cross-linked.
Since $c$ depends on $\mu_{\rm C}$, $m$  and $V_p$ (Eq.~\ref{parameter_c}), this suggests that even at the low temperature limit, the cross-linking degree of the system can still be tuned by the concentration of free linkers or the precursor density on polymer chains.
In Fig.~\ref{Fig3}ab, we plot $f_{\rm P_2C}$ and $x_p$ as functions of $\beta \Delta G$ measured in simulations of systems
with various $\mu_{\rm C}$ at $m = 5$, $n = 100$ and $\beta \mu_{\rm B} = -3$, which agree quantitatively with the theoretical prediction. One can see that both $f_{\rm P_2C}$ and $x_p$ increase and reach a plateau at low $\beta \Delta G$. In Fig.~\ref{Fig3}c, we plot the cross-linking degree at the low temperature limit  $f_{\rm P_2C}^{\infty}$  as a function of $\beta \mu_{\rm C}$. We find that $f_{\rm P_2C}^{\infty}$ decreases with increasing $\beta \mu_{\rm C}$ or decreasing $m$, which quantitatively agrees with Eq.~\ref{finf}. The physical explanation is that at $\beta \Delta G \rightarrow -\infty$, all reactive sites form either dangling $\rm PC$ bonds or cross-linking $\rm P_2C$ bonds. Breaking a cross-linking $\rm P_2C$ bond into two dangling $\rm PC$ bonds does not change the energy of the system but increases the conformational entropy of the polymers by $\Delta S_{\rm conf}$. The breaking of a $\rm P_2C$ bond also simultaneously absorbs one free cross-linker from the solution, and the translational entropy of the free cross-linker drops by $\Delta S_{\rm trans} \simeq -k_B \ln\rho_{\rm C}\Lambda^3 \simeq -\mu_{\rm C}/T$. With increasing $\mu_{\rm C}$, $\Delta S_{\rm trans}$ decreases, which favors more $\rm P_2C$ bonds breaking into dangling $\rm PC$ bonds. Similarly, at the fixed polymer length $n$ and $\phi_p$, increasing $m$ decreases $\Delta S_{\rm conf}$, which drives the formation of more cross-linking $\rm P_2C$ bonds.
Similar entropic effects in linker-mediated self-assembly were recently  observed in linker-mediated DNA-coated colloids~\cite{xia2020linker}.
In Fig.~\ref{Fig3}d, we plot the resulting saturated value of percolation parameter $x_p^{\infty}$ at $\beta \Delta G \rightarrow -\infty$ as a function of $\mu_{\rm C}$ for various $m$. One can see that $x_p^{\infty}$ also increases with decreasing $\mu_{\rm C}$, but changes more sharply than $f_{\rm P_2C}^{\infty}$. This suggests a practical way to control the mechanical property of the vitrimer at low temperature by tuning the cross-linker concentration.

\section*{Discussion and Conclusion}
In conclusion, we have proposed a  theoretical framework to { understand the cross-linking in} a newly developed linker-mediated vitrimer system. { Quantitative agreements have been found between our theoretical prediction and coarse-grained computer simulations, while the purpose of this work is to reveal the essential physics in this system, rather than seeking a quantitative prediction for real vitrimer systems.  Importantly, we find that entropy plays a determining role in these systems which is reflected in the following three aspects. }
First, we find that the  density heterogeneity of reactive sites can result in the mismatch of reaction constants of the same metathesis reaction in  the first and second  cross-linking steps, which offers a possible mechanism to explain the recent experiment~\cite{wu2020relationship}.
This mechanism of heterogeneity-enhanced reactions rates also has implications in biochemical reactions in molecular crowding conditions, which has been utilized by cells in form of membraneless compartment to enhance RNA transcription and protein translations~\cite{biochemreact}.
A similar effect of self-concentration was also reported in polymer glasses~\cite{selfcon}.
Second, we find that increasing the cross-linker concentration can induce a reentrant gel-sol transition of the vitrimer.
This offers a way to control over the reshaping or recycling of vitrimers without heating up and cooling down the system.
Lastly, in the low temperature limit, we find that the cross-linking degree of the system can still be effectively tuned by the concentration of cross-linkers, which suggests that in experiments, one should be careful with the cross-linker concentration, and lowering the temperature does not always produce highly cross-linked polymer networks. Therefore, our theory  provides  important guidelines in controlling the mechanical properties of the novel functional materials.  { It should be noted that in real vitrimer systems as studied in \cite{wu2020relationship,rottger2017high}, other factors may also influence the cross-linking in the system, e.g., the polydisperse distribution of the reactive sites on the backbone chains, the polydispersity and the rigidity of the chains themselves, the van der Waals attraction between different chemical species etc. 
Incorporating these effects within our theoretical framework is necessary to have a closer comparison between the theoretical prediction and experimental data.}
%However, we believe that the free energy contribution from these effects does not strongly depend on $f_i$ ($i = \rm PC, P_2C, B$ and C), and our theory can capture the essentially physics of the linker-mediated vitrimer system.} 

\matmethods{} 
\showmatmethods{}
\subsection*{Simulation details}
{ Simulation are initialized with 100 polymer chains in random positions. Periodic boundary conditions are applied in all three directions.
We perform grand canonical ($\mu_{\rm{B}}\mu_{\rm{C}}N_{{poly}}VT$) Monte Carlo simulations with the straight event-chain algorithm~\cite{bernard2009event,kampmann2015monte} to simulate the coarse-grained system, in which the length of each event chain is fixed at $L_c=0.2V^{1/3}$ for the translational moves of particles. 
Molecules B and C are added or deleted by using the conventional grand canonical Monte Carlo method.
Besides, we devise a bond swap algorithm to simulate reversible bond swap metathesis reactions in the vitrimer system (see below). The ratio of the trial moves, i.e., translational move, adding/removing molecules and bond swap, is $4:1:5$. In each simulation, we use $10^5$ MC moves \emph{per} particle for equilibration and $10^5$ MC moves \emph{per} particle for sampling, where the sampling frequency is  $10$ MC moves \emph{per} particle for sampling $\phi_{\rm eq}$, $f_{\rm PC}, f_{\rm P_2C}$ and $K_2/K_1$,  and $10^2$ MC moves \emph{per} particle for sampling the maximum cluster size.  
}
\subsection*{Bond swap algorithm in Event-Chain Monte Carlo simulations}
To simulate reversible bond swap reactions in vitrimer systems,  we assume that the metathesis reactions consist of two steps {as shown in Fig. 1a}, the first step reaction is
\begin{center}
\schemestart
\chemfig{\mathbf{a}_{G_0}(-[:45,,,,]\mathbf{b}_{P})(-[:315,,,,dash pattern=on 1pt off 1.5pt]\mathbf{c}_{C})}
\arrow{<->}
\chemfig{\mathbf{a}_{G_1}(-[:45,,,,dash pattern=on 1pt off 1.5pt]\mathbf{b}_{B})(-[:315,,,,]\mathbf{c}_{PC})}
\schemestop\par
\end{center}
where $\mathbf{a/b/c}$ are specific particles in the system and $\mathrm{G_0/P/C/G_1/B/PC}$ are types of the grafted point bonded with an intact precursor, precursor, cross-linker, grafted point bonded with a dangling $\rm PC$ bond, byproduct molecule and precursor-cross-linker, respectively. The acceptance probability of a proposed bond swap for the forward first step reaction is:
\begin{equation}
acc\left[ \chemfig{\mathbf{a}_{G_0}(-[:45,0.8,,,]\mathbf{b}_{P})(-[:315,0.8,,,dash pattern=on 1pt off 1.5pt]\mathbf{c}_{C})}\to
                  \chemfig{\mathbf{a}_{G_1}(-[:45,0.8,,,dash pattern=on 1pt off 1.5pt]\mathbf{b}_{\rm B})(-[:315,0.8,,,]\mathbf{c}_{PC})} 
  \right]
  =\min\left[1, \frac{ \lambda(\mathbf{a}_{G_1}, \mathbf{b}_{\rm B}) }{ \lambda( \mathbf{a}_{G_0}, \mathbf{c}_{\rm C}) } \exp{(-\beta \Delta G)}\right],
\end{equation}
where $\lambda(x,y)$ is probability of finding the pair sites $(x,y)$. $\Delta G$ is the reaction energy. Similarly, The acceptance probability of a proposed bond swap for the backward first step reaction is:
\begin{equation}
acc\left[ \chemfig{\mathbf{a}_{G_1}(-[:45,0.8,,,dash pattern=on 1pt off 1.5pt]\mathbf{b}_{B})(-[:315,0.8,,,]\mathbf{c}_{PC})} \to
                  \chemfig{\mathbf{a}_{G_0}(-[:45,0.8,,,]\mathbf{b}_{P})(-[:315,0.8,,,dash pattern=on 1pt off 1.5pt]\mathbf{c}_{C})}
  \right]
  =\min\left[1, \frac{ \lambda( \mathbf{a}_{G_0}, \mathbf{c}_{\rm C})  }{ \lambda( \mathbf{a}_{G_1}, \mathbf{b}_{\rm B})} \exp{(\beta \Delta G)}\right].
\end{equation}
For the second bond switching reaction {in Fig. 1a},
\begin{center}
\schemestart
\chemfig{\mathbf{a}_{G_0}(-[:45,,,,]\mathbf{b}_{P}-[:315,,,,dash pattern=on 1pt off 1.5pt]\mathbf{c}_{G_1}?)-[:315,,,,draw=none]\mathbf{d}_{PC}?}
\arrow{<->}
\chemfig{\mathbf{a}_{G_2}(-[:315,,,,draw=none]\mathbf{d}_{B}-[:45,,,,dash pattern=on 1pt off 1.5pt]\mathbf{c}_{G_2}?)-[:45]\mathbf{b}_{P_2C}?}
%\arrow{<->}
%\chemfig{\mathbf{a}_{G_1}(-[:315,,,,]\mathbf{d}_{PC}-[:45,,,,draw=none]\mathbf{c}_{G_0}?)-[:45,,,,dash pattern=on 1pt off 1.5pt]\mathbf{b}_{P}?} 
\schemestop\par
\end{center}
where $\mathbf{a/b/c/d}$ are specific particles in the system and $\mathrm{G_2/P_2C}$ are types of
grafted point bonded with a cross-linkng $\rm P_2C$ bond and cross-linker in a $\rm P_2C$ bond. The acceptance probability of the proposed move for the forward second step reaction is:
\begin{equation}
\begin{split}
acc\left[\chemfig{\mathbf{a}_{G_0}(-[:45,0.8,,,]\mathbf{b}_{P}-[:315,0.8,,,dash pattern=on 1pt off 1.5pt]\mathbf{c}_{G_1}?)-[:315,0.8,,,draw=none]\mathbf{d}_{PC}?}  \to
                     \chemfig{\mathbf{a}_{G_2}(-[:315,0.8,,,draw=none]\mathbf{d}_{B}-[:45,0.8,,,dash pattern=on 1pt off 1.5pt]\mathbf{c}_{G_2}?)-[:45,0.8,,,]\mathbf{b}_{P_2C}?} 
  \right]
  =\\
\min\left[1, \frac{ 2\lambda(\mathbf{c}_{G_2}, \mathbf{d}_{\rm B}) }{ \lambda(\mathbf{c}_{G_1}, \mathbf{b}_{\rm P} ) } \exp{(-\beta \Delta G)}\right],
\end{split}
\end{equation}
and the acceptance probability of the proposed move for the backward second step reaction is:
\begin{equation}
\begin{split}
acc\left[\chemfig{\mathbf{a}_{G_2}(-[:315,0.8,,,draw=none]\mathbf{d}_{B}-[:45,0.8,,,dash pattern=on 1pt off 1.5pt]\mathbf{c}_{G_2}?)-[:45,0.8,,,]\mathbf{b}_{P_2C}?} 
    \to
\chemfig{\mathbf{a}_{G_0}(-[:45,0.8,,,]\mathbf{b}_{\rm P}-[:315,0.8,,,dash pattern=on 1pt off 1.5pt]\mathbf{c}_{\rm G_1}?)-[:315,0.8,,,draw=none]\mathbf{d}_{PC}?}  
    \right]
  = \\
\min\left[1, \frac{ \lambda(\mathbf{c}_{\rm G_1}, \mathbf{b}_{\rm P} )  }{2\lambda(\mathbf{c}_{\rm G_2}, \mathbf{d}_{\rm B}) } \exp{(\beta \Delta G)}\right],
\end{split}
\end{equation}
where we double the forward reaction flux to equalize the flux between the two states before and after a second step reaction. In our simulation, we first randomly select a particle, of which the type is mentioned above and then select a potential reactive particle within the distance $r_\mathrm{cut}$ as well as the specific reaction to perform the trial move depending on the types of selected particles.

\acknow{
This work has been supported in part by the Singapore Ministry of Education through the Academic Research Fund MOE2019-T2-2-010, by Nanyang Technological University Start-Up Grant (NTU-SUG: M4081781.120), by the Advanced Manufacturing and Engineering Young Individual Research Grant (A1784C0018). We thank NSCC for granting computational resources.
}
\showacknow{}

%
%\begin{SCfigure*}[\sidecaptionrelwidth][t]
%\centering
%\includegraphics[width=11.4cm,height=11.4cm]{frog}
%\caption{This caption would be placed at the side of the figure, rather than below it.}\label{fig:side}
%\end{SCfigure*}

% Bibliography
\bibliography{reference}

\begin{thebibliography}{10}

\bibitem{montarnal2011silica}
Montarnal D, Capelot M, Tournilhac F, Leibler L (2011) Silica-like malleable
  materials from permanent organic networks.
\newblock {\em Science} 334(6058):965--968.

\bibitem{capelot2012metal}
Capelot M, Montarnal D, Tournilhac F, Leibler L (2012) Metal-catalyzed
  transesterification for healing and assembling of thermosets.
\newblock {\em Journal of the american chemical society} 134(18):7664--7667.

\bibitem{smallenburg2013patchy}
Smallenburg F, Leibler L, Sciortino F (2013) Patchy particle model for
  vitrimers.
\newblock {\em Phys Rev Lett} 111(18):188002.

\bibitem{ciarella2018dynamics}
Ciarella S, Sciortino F, Ellenbroek WG (2018) Dynamics of vitrimers: Defects as
  a highway to stress relaxation.
\newblock {\em Phys Rev Lett} 121(5):058003.

\bibitem{ciarella2019understanding}
Ciarella S, Biezemans RA, Janssen LM (2019) Understanding, predicting, and
  tuning the fragility of vitrimeric polymers.
\newblock {\em Proceedings of the National Academy of Sciences}
  116(50):25013--25022.

\bibitem{meng2019elasticity}
Meng F, Saed MO, Terentjev EM (2019) Elasticity and relaxation in full and
  partial vitrimer networks.
\newblock {\em Macromolecules} 52(19):7423--7429.

\bibitem{tito2019harnessing}
Tito NB, Creton C, Storm C, Ellenbroek WG (2019) Harnessing entropy to enhance
  toughness in reversibly crosslinked polymer networks.
\newblock {\em Soft matter} 15(10):2190--2203.

\bibitem{wu2019dynamics}
Wu JB, Li SJ, Liu H, Qian HJ, Lu ZY (2019) Dynamics and reaction kinetics of
  coarse-grained bulk vitrimers: a molecular dynamics study.
\newblock {\em Phys. Chem. Chem. Phys.} 21(24):13258--13267.

\bibitem{Denissen2016}
Denissen W, Winne JM, Du~Prez FE (2016) Vitrimers: permanent organic networks
  with glass-like fluidity.
\newblock {\em Chem. Sci.} 7:30--38.

\bibitem{Winne2019}
Winne JM, Leibler L, Du~Prez FE (2019) Dynamic covalent chemistry in polymer
  networks: a mechanistic perspective.
\newblock {\em Polym. Chem.} 10:6091--6108.

\bibitem{zee2020}
Van~Zee NJ, Nicola{\"y} R (2020) Vitrimers: Permanently crosslinked polymers
  with dynamic network topology.
\newblock {\em Prog. Polym. Sci.} 104:101233.

\bibitem{zhang2018polymer}
Zhang ZP, Rong MZ, Zhang MQ (2018) Polymer engineering based on reversible
  covalent chemistry: A promising innovative pathway towards new materials and
  new functionalities.
\newblock {\em Prog. Polym. Sci.} 80:39--93.

\bibitem{brutman2014polylactide}
Brutman JP, Delgado PA, Hillmyer MA (2014) Polylactide vitrimers.
\newblock {\em ACS Macro Letters} 3(7):607--610.

\bibitem{denissen2015vinylogous}
Denissen W, et~al. (2015) Vinylogous urethane vitrimers.
\newblock {\em Advanced Functional Materials} 25(16):2451--2457.

\bibitem{taynton2014heat}
Taynton P, et~al. (2014) Heat-or water-driven malleability in a highly
  recyclable covalent network polymer.
\newblock {\em Advanced materials} 26(23):3938--3942.

\bibitem{wojtecki2011using}
Wojtecki RJ, Meador MA, Rowan SJ (2011) Using the dynamic bond to access
  macroscopically responsive structurally dynamic polymers.
\newblock {\em Nature materials} 10(1):14--27.

\bibitem{zhang2014application}
Zhang MQ, Rong MZ, , et~al. (2014) Application of alkoxyamine in self-healing
  of epoxy.
\newblock {\em Journal of Materials Chemistry A} 2(18):6558--6566.

\bibitem{lu2012making}
Lu YX, Tournilhac F, Leibler L, Guan Z (2012) Making insoluble polymer networks
  malleable via olefin metathesis.
\newblock {\em Journal of the American Chemical Society} 134(20):8424--8427.

\bibitem{pepels2013self}
Pepels M, Filot I, Klumperman B, Goossens H (2013) Self-healing systems based
  on disulfide--thiol exchange reactions.
\newblock {\em Polymer Chemistry} 4(18):4955--4965.

\bibitem{rottger2017high}
R{\"o}ttger M, et~al. (2017) High-performance vitrimers from commodity
  thermoplastics through dioxaborolane metathesis.
\newblock {\em Science} 356(6333):62--65.

\bibitem{garcia2017future}
Garcia JM, Robertson ML (2017) The future of plastics recycling.
\newblock {\em Science} 358(6365):870--872.

\bibitem{wu2020relationship}
Wu S, Yang H, Huang S, Chen Q (2020) Relationship between reaction kinetics and
  chain dynamics of vitrimers based on dioxaborolane metathesis.
\newblock {\em Macromolecules} 53(4):1180--1190.

\bibitem{nihspolymer}
Ni R, Dijkstra M (2013) Effect of bond length fluctuations on crystal
  nucleation of hard bead chains.
\newblock {\em Soft Matter} 9:365.

\bibitem{hansen1990theory}
Hansen JP, McDonald IR (1990) {\em Theory of simple liquids}.
\newblock (Elsevier).

\bibitem{ricarte2018phase}
Ricarte RG, Tournilhac F, Leibler L (2018) Phase separation and self-assembly
  in vitrimers: Hierarchical morphology of molten and semicrystalline
  polyethylene/dioxaborolane maleimide systems.
\newblock {\em Macromolecules} 52(2):432--443.

\bibitem{Ralm2020ma}
Ricarte RG, Tournilha F, Clitre M, Leibler L (2020) Linear viscoelasticity and
  flow of self-assembled vitrimers: The case of a polyethylene/dioxaborolane
  system.
\newblock {\em Macromolecules} 53(5):1852--1866.

\bibitem{tanaka2011polymer}
Tanaka F (2011) {\em Polymer physics: applications to molecular association and
  thermoreversible gelation}.
\newblock (Cambridge University Press).

\bibitem{xia2020linker}
Xia X, Hu H, Ciamarra MP, Ni R (2020) Linker-mediated self-assembly of mobile
  dna-coated colloids.
\newblock {\em Sci. Adv.} 6(21):eaaz6921.

\bibitem{biochemreact}
Shin Y, Brangwynne CP (2017) Liquid phase condensation in cell physiology and
  disease.
\newblock {\em Science} 357(6357):eaaf4382.

\bibitem{selfcon}
Lodge TP, McLeish TCB (2000) Self-concentrations and effective glass transition
  temperatures in polymer blends.
\newblock {\em Macromolecules} 33:5278.

\bibitem{bernard2009event}
Bernard EP, Krauth W, Wilson DB (2009) Event-chain monte carlo algorithms for
  hard-sphere systems.
\newblock {\em Phys. Rev. E} 80(5):056704.

\bibitem{kampmann2015monte}
Kampmann TA, Boltz HH, Kierfeld J (2015) Monte carlo simulation of dense
  polymer melts using event chain algorithms.
\newblock {\em The Journal of chemical physics} 143(4):044105.

\end{thebibliography}

\end{document}